\documentclass[10pt,epsf]{article}
\usepackage{hyperref}
\usepackage{graphics}
\usepackage{amssymb}
\usepackage{epsf}
\usepackage{epsfig}

\def\ds#1{#1\kern-1ex\hbox{/}}
\def\dsh{h\kern-1.2ex /}

\newcommand{\bea}{\begin{eqnarray}}
\newcommand{\eea}{\end{eqnarray}}

\def\beq{\begin{equation}}
\def\eeq{\end{equation}}

\def\beqn{\begin{eqnarray}}
\def\eeqn{\end{eqnarray}}
\def\ba{\begin{eqnarray}}
\def\ea{\end{eqnarray}}

\newcommand{\beqa}{\begin{eqnarray}}
\newcommand{\eeqa}{\end{eqnarray}}

\begin{document}
\begin{center}
\vspace{1.cm}
{\bf \large $Z^{\prime}$ Searches at the LHC: Some QCD Precision Studies in Drell-Yan\\ }
\vspace{1.5cm}
{\bf $^{a}$Claudio Corian\`{o} $^b$Alon E. Faraggi $^{a}$Marco Guzzi}

\vspace{1cm}

{\it $^a$Dipartimento di Fisica, Universit\`{a} del Salento \\
and  INFN Sezione di Lecce,  Via Arnesano 73100 Lecce, Italy}\\
\vspace{.5cm}
{\it $^b$ Department of Mathematical Sciences,
              University of Liverpool, Liverpool L69 7ZL, United Kingdom
				                            }\\
\vspace{.5cm}

\begin{abstract}
Discovery potentials for extra neutral interactions at the Large Hadron Collider
in forthcoming experiments are analyzed in Drell-Yan. For this
purpose we use high precision next-to-next-to-leading order (NNLO) determination
of the invariant mass distributions and of the total cross sections
in the kinematic region around 1 TeV.
In this region we explore the possibility to make a preliminary distinction between
different anomaly-free extensions of the Standard Model.
\end{abstract}
\end{center}
\newpage

\section{Introduction}
Searching for extra neutral interactions at the Large
Hadron Collider involves a combined effort from two
sides: precise determination of the signal, which should
allow a discrimination of any specific model, and
precise determination of the SM background, which is
a very difficult task at a hadron collider due to the
presence of the QCD effects in hadronization.
Extra $Z^{\prime}$ come from different extensions of the
Standard Model like in
left-right symmetric models, in Grand Unified Theories 
(GUTs) and in string inspired constructions.
It has also been suggested that
the existence of a low scale $Z^{\prime}$ may
account for the suppression of proton decay mediating
operators (free fermionic models) in supersymmetric theories and otherwise
\cite{Coriano:2007ba} \cite{Coriano:2008wf}. Some of these U(1) could also be anomalous,
and invoke a mechanism of cancelation of the anomalies
that requires an axion \cite{Coriano':2005js} \cite{Coriano:2007fw}
\cite{Coriano:2007xg} \cite{Armillis:2007tb} \cite{Coriano:2008pg} \cite{Anastasopoulos:2008jt}.

\section{The interaction Lagrangian}

In our analysis we have decided to compare our results for a string-inspired $Z^{\prime}$ with a series of
models studied in \cite{Carena:2004xs}. We just mention that the
construction of models with extra $Z^\prime$ using a bottom-up approach
is, in general, rather straightforward, being based mostly
on the principle of cancelation of the gauge 
cubic $U(1)_{Z^\prime}^3$ and the mixed and gravitational anomalies.

After the diagonalization of the mass matrix we obtain the physical
masses of the gauge bosons
\begin{equation}
M_Z^2=\frac{g^2}{4
\cos^2\theta_W}(v_{H_1}^2+v_{H_2}^2)\left[1+O(\varepsilon^2)\right];
\hspace{0.4cm}
M_{Z^{\prime}}^2=\frac{g_z^2}{4}(z_{H_1}^2
v_{H_1}^2+z_{H_2}^2v_{H_2}^2+z_{\phi}^2
v_{\phi}^2)\left[1+O(\varepsilon^2)\right].
\end{equation}
where $\varepsilon$ is defined as a mixing-perturbative parameter $(\varepsilon\approx 10^{-3})$ and
$v_{H_i}, v_{\phi}$  are the expectation values of the two Higgses of the theory, while $z_{H_i}, z_{\phi}$ denote their charges. 

The colour-averaged inclusive differential cross section
for the reaction $P + P \rightarrow l_1 +l_2 +X $, is given by
\ba
\frac{d\sigma}{dQ^2}=\tau \sigma_{V}(Q^2,M_V^2)
W_{V}(\tau,Q^2)\hspace{1cm} \tau=\frac{Q^2}{S},
\ea
where all the hadronic initial state information is contained in the
hadronic structure function $W_{V}(\tau,Q^2)$ which is defined in \cite{Hamberg:1990np}
while all the information regarding the $Z^{\prime}$ interactions with
the quarks and the leptons is contained in the point like cross section
defined as
\ba
\sigma_{{Z^{\prime}}}(Q^2)=\frac{\pi\alpha_{em}}{4
M_{{Z^{\prime}}}\sin^2\theta_W \cos^2\theta_W N_c}
\frac{\Gamma_{{Z^{\prime}}\rightarrow \bar{l}
l}}{(Q^2-M_{Z^{\prime}}^2)^2 + M_{Z^{\prime}}^2 \Gamma_{Z^{\prime}}^2}.
\ea
Here $\Gamma_{Z^{\prime}}$ and $\Gamma_{{Z^{\prime}}\rightarrow \bar{l} l}$
are respectively the total and the partial decay rate of the $Z^{\prime}$.

Our results for the NLO total cross sections are listed in Tab.~\ref{TabNLO12}.

\begin{table}
\begin{center}
\begin{footnotesize}
\begin{tabular}{|c||c|c|c|c|c|}
\hline
\multicolumn{6}{|c|}{$\sigma_{tot}^{nlo}$ [fb], $\sqrt{S}=14$ TeV, $M_{Z^{\prime}}=1.2$ TeV, $\tan\beta=40$, Candia evol.}
\tabularnewline
\hline
$g_z$&
Free Ferm.  &
$U(1)_{B-L}$&
$U(1)_{q+u}$&
$U(1)_{10+\bar{5}}$&
$U(1)_{d-u}$
\tabularnewline
\hline
\hline
$0.1$ & $0.572$& $1.620$& $1.224$& $0.367$ & $0.309$ \\
       & $0.146$& $0.160$& $0.213$& $0.079$ & $0.027$ \\
       & $0.008$& $0.202$& $0.114$& $0.010$ & $0.013$
\tabularnewline
\hline
$0.3$ & $5.418$& $15.559$& $12.412$& $3.281$ & $2.427$ \\
       & $1.314$& $1.439$& $1.916$& $0.715$ & $0.240$ \\
       & $0.074$& $1.936$& $1.160$& $0.091$ & $0.101$
\tabularnewline
\hline
$0.5$ & $14.316$& $40.465$& $30.535$& $9.149$ & $6.741$ \\
       & $3.650$& $3.997$& $5.323$& $1.987$ & $0.667$ \\
       & $0.195$& $5.036$& $2.853$& $0.255$ & $0.279$
\tabularnewline
\hline
$0.7$ & $28.077$& $79.270$& $59.836$& $17.915$ & $13.212$ \\
       & $7.154$& $7.833$& $10.433$& $3.894$ & $1.307$ \\
       & $0.383$& $9.865$& $5.591$& $0.498$ & $0.547$
\tabularnewline
\hline
$1.0$ & $57.394$& $161.625$& $121.921$& $36.556$ & $26.959$ \\
       & $14.600$& $15.986$& $21.292$& $7.946$ & $2.667$ \\
       & $0.783$& $20.114$& $11.392$& $1.017$ & $1.117$
\tabularnewline
\hline
\end{tabular}
\end{footnotesize}
\end{center}
\caption{Total cross sections at NLO, $M_{Z^{\prime}}=1.2$ TeV.  }
\label{TabNLO12}
\end{table}

\section{Discussion}

While the analysis of the invariant mass distributions shows in general an overlap
between the various models in both the NLO and NNLO cases that we have studied \cite{Coriano:2008wf}, the differences among the predictions seem to be different if we look at observables
like the total cross section, which is obtained by integrating the invariant mass
distributions over $\pm 3\Gamma_{Z^{\prime}}$ around the peak. This feature
depends on the fact that in this interval all the models exhibit
differences in their shape, width and in their peak values.
Therefore, by focusing the attention near the resonance one can hope to better distinguish
among the various models.
\section{Conclusions}

The possibility of discovering extra $Z^{\prime}$ at the LHC is
realistic, being they common both in GUTs and string-inspired
models. Therefore, precision determinations of the QCD background are necessary to identify them at the
LHC. Some golden-plated processes like
$Z \rightarrow \gamma \gamma$ or $Z \rightarrow l\bar{l}l\bar{l}$
and Drell-Yan are the best place to search for new signals. From the string theory side, 
the family of free fermionic models is
one of the most phenomenologically attractive, because it addresses quite successfully the issue of proton stability \cite{Coriano:2007ba}.  

Since the V-A structure of the couplings is different in each model, a measurement  of
forward-backward asymmetries and/or of charge asymmetries
could be helpful \cite{Petriello:2008zr} for their detection, but this can happen 
only if the gauge coupling is sizeable.
However, discriminating among the models remains a difficult issue for which NNLO QCD determinations, at
least in leptoproduction, though useful, do not seem to be necessary in
the immediate future - for such large values of the mass of the extra $Z^\prime$ -,
while the NLO effects remain important for the reduction of the renormalization/factorization scale 
dependence of the cross sections.


\section*{Acknowledgments}
M.G. would like to thank the I.N.F.N. group of Bologna for the financial support.
A great thank goes also to Gennaro Corcella, Barbara Simoni and Vincenzo Vagnoni
for all the support.

\bibliographystyle{h-elsevier3}
\bibliography{ifae2}
\end{document}